\documentstyle{elsart}
\begin {document}

\begin {frontmatter}
\title {Effective action for Dirac spinors in the presence \\
    of general uniform electromagnetic fields}

\author [Bologna]{Roberto Soldati\thanksref{mailr}},
\author [Trieste]{Lorenzo Sorbo\thanksref{maill}}

\address[Bologna]{Dipartimento di Fisica "A. Righi"\\
          via Irnerio 46 -- 40126 Bologna, Italia}
\address[Trieste]{S.I.S.S.A.\\
          via Beirut 2-4 -- 34013 Trieste, Italia}

\thanks[mailr]{E-mail address: soldati@bo.infn.it}
\thanks[maill]{E-mail address: sorbo@sissa.it}

\begin{abstract}
Some new expressions are found, concerning the one-loop effective action
of four dimensional massive and massless Dirac fermions in the presence
of general uniform electric and magnetic fields, with
$\vec E\cdot \vec H\neq 0$ and $\vec E ^2\neq \vec H ^2$. 
The rate of pair-production
is computed and briefly discussed.
\end{abstract}
\end{frontmatter}

Non-renormalizable effective field theories turn out to be quite useful in
the description of physics below some specific momentum scale. One of the early
known examples is provided, taking one-loop corrections into account, 
by the (nonlinear) 
effective lagrangean for the electromagnetic field in the presence of virtual
fermions, which enables to describe light-light scattering at low momenta of
the order of the electron mass.
 As it is well known, the use of path-integral techniques allows 
to obtain the correction to the classical action in terms of the evaluation 
of the determinant of the Dirac operator. The first efforts in this direction
date back to Euler and Heisenberg \cite {EH}, who worked out an implicit 
expression of the effective Lagrange function for Maxwell theory in the 
context of electron-hole theory. Later on, Schwinger \cite {Sch} derived a 
gauge-invariant integral representation of the effective lagrangean by means of
the so-called "proper-time" technique. Then, during the next three-four 
decades, no step was made towards an explicit expression of an effective theory 
of electromagnetism. The introduction of Hawking's $\zeta$-function technique
\cite {Haw} renewed the interest in the subject and some further progress in 
this direction was put forward: Blau, Visser and Wipf 
\cite {blau} obtained an analytic form for the effective action of a uniform 
electromagnetic field in any number of space-time dimensions, both for massive 
and for massless fermions.

Unfortunately, none of these expressions is fully satisfactory: on the one hand,
Schwinger's one, besides being implicit, is not valid in a theory with 
massless fermions. On the other hand, the expressions given in ref.\cite{blau}
are explicit and valid also for massless Fermi particles, but they are 
established
only for some particular configurations of the uniform electric and magnetic 
fields. In this paper a completion will be given to the work of Blau 
{\it et al.}, in order to obtain a general and explicit expression for the 
effective lagrangean of massive and massless QED in the presence of uniform 
general electromagnetic fields.
 
The use of the path-integral method forces us to work in the euclidean 
framework: only at the very end of our calculations we shall
operate a Wick rotation to Minkowski space-time. The effective euclidean 
action, in the one-loop approximation, is given by
\begin {equation}
{\cal S}^{\mathrm E}_{\mathrm {Eff}}[A_\mu ]={\cal S}^{\mathrm E}_{
\mathrm {Cl}}[A_\mu ]-\log {\mathrm {Det}}\bigl (\,
{\FMslash D} [A_\mu ]+im\bigr )
\end {equation}
where ${\cal S}_{\mathrm {Cl}}^{\mathrm E} [A_\mu ]$ is the classical 
euclidean action $\int \d^4 x F_{\mu\nu} F_{\mu\nu}/4$ and 
${\FMslash D} [A_\mu ]\equiv \gamma_\mu (\partial_\mu -ieA_\mu )$ is the 
euclidean Dirac operator, $m$ being the fermion mass. As the Dirac 
operator is normal ($[\,{\FMslash D} +im,{\FMslash D}^\dagger -im]=0$), we have
\begin {equation}
\bigl |{\mathrm {Det}}({\FMslash D}+im)\bigr |^2={\mathrm {Det}}
({\FMslash D}^\dagger{\FMslash D}+m^2)
\end{equation}
and consequently the effective lagrangean is defined (by means of the 
$\zeta$-function regularization) to be
\begin {equation}
{\cal S}^{\mathrm E}_{\mathrm {Eff}}[A_\mu ]={\cal S}^{\mathrm E}_{
\mathrm {Cl}}[A_\mu ]-\frac {1}{2} \frac {\partial}{\partial s}\zeta\bigl (
s;[A_\mu]\bigr )\Bigr |_{s=0} 
\end{equation}

Since the square of the Dirac operator 
${\FMslash D}^\dagger{\FMslash D}=(\partial_\mu -ieA_\mu)^2-eF_{\mu\nu}\Sigma_{\mu\nu}/2$ 
is elliptic, its determinant 
can be evaluated by means of the $\zeta$-function regularization (the matrices
$\Sigma_{\mu\nu}$ are defined as $\Sigma_{\mu\nu}\equiv {\mathrm i}
[\gamma_\mu ,\gamma_\nu ]/2$): to perform this evaluation, it is necessary 
to obtain the spectrum of the operator, which can be explicitly calculated in 
the case of uniform fields. In this particular situation, a frame can always 
be chosen such that $F_{03}=-F_{30}=E$, $F_{12}=-F_{21}=B$, all the other 
components of the field-strength tensor vanishing: the spectrum of the operator
${\FMslash D}^\dagger{\FMslash D}$ turns out to be of the form
\begin {eqnarray}
\lambda_{n_E,n_B}=2|eE|n_E&+&2|eB|n_B\\ 
 &(&\hskip -1.5pt n_E,n_B=0,1,2,... )\nonumber
\end{eqnarray}
(see, for instance, \cite {bas}).

The first fact we can notice is that $E$ and $B$ play a symmetric role in 
euclidean space (this will not be the case in Minkowski space-time). Taking
into account the degeneracy of each eigenvalue (which can be obtaind by
Landau levels counting), we get, for the $\zeta$-function, the following
implicit expression: namely,
\begin {eqnarray}
&\zeta&(s)=(vol)_4 \frac{e^2|EB|}{4\pi^2}\Biggl [\,4
\sum_{n_E=1}^\infty\sum_{n_B=1}^\infty\biggl (\frac{2|eE|n_E+2|eB|n_B+m^2}
{\mu^2}\biggr )^{-s}+\nonumber\\ 
&+&\biggl (\frac {m^2}{\mu^2}\biggr )^{-s}+2\sum_{n_E=1}^\infty\biggl (
\frac {2|eE|n_E+m^2}{\mu^2}\biggr )^{-s}+2\sum_{n_B=1}^\infty\biggl (
\frac {2|eB|n_B+m^2}{\mu^2}\biggr )^{-s}\Biggr ]
\label {zdef}
\end {eqnarray}
where $(vol)_4$ is the (regularized) volume of the four-dimensional euclidean
space and $\mu$ is a normalization parameter with dimensions of a mass. The 
$\zeta$-function defined in eq.(\ref {zdef}) is singular in the massless limit, 
owing to the presence of zero-modes for the squared massless Dirac operator 
${\FMslash D}^\dagger{\FMslash D}$; this limit will be further examinated below.\par
The spectrum of the squared Dirac operator turns out to be quite different
when either $E$ or $B$ vanish. Taking, for instance, $B=0$, the
spectrum will be of the form
\begin {eqnarray}
\lambda_{n_E,p_1,p_2}=2|eE|n_E&+&p_1^2+p_2^2 \\
&(&p_1,p_2\in R ,n_E=0,1,2...)\nonumber
\end {eqnarray}
Obviously, if $B=0$ the $\zeta$-function will have a different expression
with respect to eq.(\ref{zdef}).
Here we are not interested in this expression, because the $\zeta$-function,
in the case of vanishing electric -- or magnetic -- field, may be also 
calculated exploiting the (well known) 
asymptotic behavior of the heat kernel, as is done
in ref. \cite {blau}. There, 
the authors worked out explicitly the effective
action (both in theories with $m\neq 0$ and $m=0$) in the case of vanishing
$E$ (or $B$) and also, from expression (\ref {zdef}), in the case $|E|=|B|$. In 
what follows, we will obtain an asymptotic expansion of the $\zeta$-function 
in powers of $|E/B|$.

In order to simplify the notation, we define the dimensionless quantities
$a\equiv |eE|/\mu^2$, $b\equiv |eB|/\mu^2$ and $c\equiv m^2/\mu^2$: the
$\zeta$-function may be cast in integral form as
\begin {equation}
\zeta (s;a,b;c)=\mu^4\frac{(vol)_4}{4 \pi^2}\frac {ab}{\Gamma (s)}
\int_0^\infty \d t\ t^{s-1} {\e}^{-ct}\biggl [ 2 \frac {{\e}^{-2at}
+{\e}^{-2bt}}{(1-{\e}^{-2at})(1-{\e}^{-2bt})}+1\biggr ]
\label {zintm:com}
\end {equation}
\par
One of the main difficulties in the analysis of the $\zeta$-function for the
four-dimensional Dirac theory comes from the fact that it is a function of the
two arguments $a$ and $b$ (for some fixed values of the parameters $s$ and 
$c$). This difficulty can be overcome by noticing that the $\zeta$-function 
(\ref{zintm:com}) obeys the scaling property
\begin {equation}
\zeta (s;a,b;c)=a^{2-s}\zeta \Bigl (s;1,\frac{b}{a};\frac{c}{a}\Bigr )
\label {scalm}
\end {equation}
so that, in order to find the general expression of the effective action, it 
is sufficient to study the function of the single ratio $b/a\equiv y$
\begin {equation}
g(s;y,z)\equiv \frac{y}{\Gamma (s)}\int_0^\infty \d t\ t^{s-1} {\e}^{-zt}
\biggl [ 2 \frac{\e^{-2t}+\e^{-2yt}}{(1-\e^{-2t})(1-\e^{-2yt})}+1\biggr ]
\label {zintm:red}
\end {equation}
where $z\equiv c/a$.
The explicit formulae reported by Blau {\it et al.} are valid for $E=0$ and 
$|E|=|B|$, namely $y=0$ and $y=1$. The expansion around $y=1$ can be performed
quite easily, and is left to the interested reader. We aim now to investigate
the behavior of $g(s;y,z)$ when $y\simeq 0$: after some simple manipulations
we can rewrite the function $g(s;y,z)$ as
\begin {equation}
g(s;y,z)=\frac {2^{-s}y}{\Gamma (s)}\int_0^\infty \d t\ t^{s-1}\e^{-\frac{z}{2}
t}\frac {1+\e^{-t}}{1-\e^{-t}}\coth\Bigl (\frac {yt}{2}\Bigr )
\label {zintm:red2}
\end {equation}
\par
In order to expand the function $\coth (yt/2)$ in a power series of  
$yt/2$, it is necessary to cut the integral in eq.(\ref {zintm:red2}) at the 
upper end: we restrict the integration domain to the interval $[0,y^{-\alpha}]$
(where $0<\alpha<1$). With such a choice, in the limit $y\rightarrow 0$ the 
domain of the $t$ variable extends to the whole set of the real positive 
numbers. On the other hand, the quantity $yt/2<y^{1-\alpha}/2$ approaches zero
in this limit, so that we can expand $\coth(yt/2)$ in power series of its 
argument. We must only assure ourselves that the error made when cutting the 
upper integration limit does itself approach zero when $y\rightarrow 0$: it
can be shown that this is what really happens, since
\begin {equation}
\frac {y}{\Gamma (s)}\int_{y^{-\alpha}}^\infty \d t\ t^{s-1}\e^{-\frac {z}{2}t}
\frac {1+\e^{-t}}{1-\e^{-t}}\coth\Bigl (\frac {yt}{2}\Bigr )<
\e^{-\frac{z}{2}y^{-\alpha}}f_1(y)
\label {est1}
\end{equation}
where $f_1(y)$ diverges at most as a power of $1/y$ when $y\rightarrow 0$ so
that, as long as $z\neq 0$, the error made by cutting the integral is 
exponentially small when $y\rightarrow 0$.\par
Now we can safely expand the function $g(s;y,z)$ in power series of $y$: the 
estimate (10) allows us to write 
\begin {eqnarray}
g(s;y,z)=\frac{2^{1-s}}{\Gamma (s)}\int_0^{y^{-\alpha}} \d t\ t^{s-2}
&\e&^{-\frac {z}{2} t}\frac {1+\e^{-t}}{1-\e^{-t}}\Bigl [1+\sum_{k=1}^N
\frac {{B}_{2k}}{(2k)!} \bigl (yt \bigr )^{2k}\Bigr ] \nonumber\\
&+&O\bigl (y^{2N+2}\bigr )+\e^{-\frac {z}{2} y^{-\alpha}}f_1 (y)
\label {zintm:pow}
\end {eqnarray}
where ${B}_{2k}$ is the $2k$-th Bernoulli number. This expression, in 
the limit $y\rightarrow 0$, yields an asymptotic series in $y$, whose 
coefficients can be all explicitly computed: namely,
\begin {eqnarray}
g(s;y,z)&=&\frac {2^{1-s}}{\Gamma (s)}\sum_{k=0}^N\frac {{\bf B}_{2k}}
{(2k)!}y^{2k} \Gamma (s-1+2k)\cdot\nonumber\\ &\cdot&\biggl [
2\zeta_H \Bigl (s-1+2k;1+\frac {z}{2}\Bigr )+\Bigl (\frac {z}{2}\Bigr )^{1-2k-s}
\biggr ]+O\bigl (y^{2N+2}\bigr )
\label {zpowm}
\end {eqnarray}
$\zeta_H(a;x)\equiv \sum_{n=0}^\infty (n+x)^{-a}$ being the Hurwitz 
$\zeta$-function. Exploiting the scaling property (\ref{scalm}), the effective 
euclidean lagrangean of QED in the presence of uniform fields can be easily 
calculated as a power series of $|B/E|$ (or $|E/B|$, owing to the symmetry 
between 
the electric and magnetic field in the euclidean space).\par
The massless limit unravels a quite different behavior: this fact is a 
consequence of the explicit exclusion of the zero-modes from the 
$\zeta$-function (\ref {zdef}), whose integral representation will be, in this 
case, of the form
\begin {equation}
\zeta (s;a,b)=\mu^4\frac {(vol)_4}{4\pi^2} 2 \frac {ab}{\Gamma (s)}\int_0
^\infty \d t\ t^{s-1}\frac {\e^{-2at}+\e^{-2bt}}{(1-\e^{-2at})(1-\e^{-2bt})}
\label {zintn:com}
\end{equation}
\par
The scaling property $\zeta (s;a,b)=a^{2-s}\zeta (s;1,b/a)$ allows us to
extract all the information we need by the analysis of a function of only 
one variable, namely
\begin {equation}
g(s;y)=\frac {2^{1-s}y}{\Gamma (s)}\int_0^\infty \d t\ t^{s-1}\frac {\e^{-t}+
\e^{-yt}}{(1-\e^{-t})(1-\e^{-yt})}
\label{zintn:red}
\end{equation}
\par
Now we would like to proceed as in the massive ($z\neq 0$) theory, but it 
is apparent from the estimate (\ref {est1}) that, for vanishing $z$, the 
error made by cutting the integral at $y^{-\alpha}$ at the upper end is not 
negligible in the limit of small $y$. Yet, it can be easily verified that
\begin{equation}
g(s;y)-(2y)^{1-s}\zeta_R (s)=\frac {2^{1-s}y}{\Gamma (s)}\int_0^\infty
\d t\ t^{s-1}\frac {\e^{-t}}{1-\e^{-t}}\coth\Bigl (\frac {yt}{2}\Bigr )
\label {zintn:red2}
\end{equation}
where $\zeta_R(a)\equiv\zeta_H(a;1)$ is the Riemann $\zeta$-function. Cutting 
the integral of eq.(\ref {zintn:red2}) at the upper end we get the estimate
\begin{equation}
\int_{y^{-\alpha}}^\infty \d t\ t^{s-1}\frac {\e^{-t}}{1-\e^{-t}}\coth
\Bigl (\frac {yt}{2}\Bigr )<\e^{-y^{-\alpha}} f_2(y)
\label {est2}
\end {equation}
where $f_2(y)$ has a behavior analogous to the one of $f_1(y)$. Now we are 
allowed to expand $g(s;y)-(2y)^{1-s}\zeta_R(y)$ in power series of $y$,
obtaining
\begin {eqnarray} 
g(s;y)=(2y)^{1-s}\zeta_R(s)+\frac {2^{2-s}}{\Gamma (s)}\sum_{k=0}^N
\frac {{\bf B}_{2k}}{(2k)!}\Gamma (s-1&+&2k)\zeta_R(s-1+2k) y^{2k}+\nonumber\\
&+&O\bigl (y^{2N+2}\bigr )
\label {zpown}
\end {eqnarray}
\par
We have performed all of the calculations necessary to gain an expression of
the effective lagrangean for massive and massless QED in an asymptotic series 
of $|E/B|$: now we will show how these corrections to the $E=0$ case work, paying 
special attention to the rate of production of fermion-antifermion pairs
and to the behavior under parity of the massless effective theory. To this aim, 
it will be necessary to obtain the effective lagrangean in minkowskian
space-time. First of all, we will analyse the "unperturbed" problem ($E=0$ or,
equivalently, $B=0$): in this limit the effective lagrangean can be found in 
\cite {blau}, but in that work the transition to the Lorentz metric is not 
considered. Denoting by $\cal E$ and $\cal B$ the minkowskian electric
and magnetic field strengths respectively
(whereas in euclidean space we used the symbols
$E$ and $B$), the transition to the Minkowski space-time is performed by 
means of the substitution
\begin {equation}
{\cal L}_{\mathrm {Eff}}^{\mathrm M}({\cal E},{\cal B})=
-{\cal L}_{\mathrm {Eff}}^{\mathrm E}(E=-{\mathrm i}{\cal E},
B={\cal B})
\label {wickEB}
\end{equation}
where the superscripts $\mathrm E$ and $\mathrm M$ mean, obviously, euclidean 
and minkowskian metric.\par
If ${\cal E}=0$ the effective theory in Minkowski space
does not give any new information, being trivially 
${\cal L}_{\mathrm {Eff}}^{\mathrm M}({\cal B})=
-{\cal L}_{\mathrm {Eff}}^{\mathrm E}(B={\cal B})$. It is much more 
interesting to analyse the case 
${\cal B}=0$, since, from eq.(5.1.8) of ref.\cite{blau}, the effective 
lagrangean is obtained to be
\begin {eqnarray}
{\cal L}_{\mathrm {Eff}}^{\mathrm M}({\cal E},0)=\frac {{\cal E}^2}{2}-
\frac{e^2{\cal E}^2}{2\pi^2}&\biggl \{&
\biggl [1-\log \Bigl (-\frac {2{\mathrm i} e{\cal E}}{\mu^2}\Bigr )\biggr ]
\zeta_H\Bigl (-1;1+{\mathrm i}\frac {m^2}{2e{\cal E}}\Bigr )+\nonumber \\
&+&\zeta_H'\Bigl (-1;1+{\mathrm i}\frac {m^2}{2e{\cal E}}\Bigr )\biggr \}+
{\mathrm i}\frac {m^2 e{\cal E}}
{8\pi^2}\Bigl [\log \frac {m^2}{\mu^2}-1\Bigr ]
\label {leff:b0}
\end {eqnarray}
and turns out to have a nonvanishing imaginary part. The imaginary 
(absorbitive) part is interpreted as half of the probability of generation of 
fermion-antifermion pairs per unit of space-time (see also \cite{IZ}, 
chapt. 4-3). The value of this probability cannot be easily obtained from 
eq.(\ref {leff:b0}), because it is quite 
hard to evaluate the real and the imaginary 
part of the Hurwitz $\zeta$ function of complex argument. On the other hand, in
\cite {blau} both the strong ($E\gg m^2$) and the weak field limits of the 
effective lagrangean are exhibited in a form which does not contain the Hurwitz
special function: so, after having performed the transition to 
the minkowskian space-time, we can write explicitly the strong- and 
weak- field approximations of the rate of pair creation in a purely 
electric uniform  field. The exact result is given (in implicit form) in ref.
\cite{IZ}, so that we can recover the expressions given by
Blau {\it et al.} and check the validity of eq.(\ref {wickEB}) we used to 
"Wick-rotate" the electromagnetic field: the exact pair production rate per 
unit space-time is (\cite{IZ}, par. 4-3-3)
\begin {equation}
w=\frac {e^2{\cal E}^2}{4\pi^3}\sum_{n=1}^\infty\frac {1}{n^2}\exp \Bigl \{-n
\frac {\pi m^2}{e{\cal E}}\Bigr \}
\label {rate:exact}
\end{equation}
\par
From the expressions of \cite{blau} we have that the weak-field expansion 
gives $w=0$: this fact is in agreement with eq.(\ref{rate:exact}), which shows 
that the function $w({\cal E})$ is not analytic in ${\cal E}=0$. On the other 
hand, for strong fields we get 
\begin{equation}
w=\frac {e^2{\cal E}^2}{24 \pi}-\frac {m^2 e{\cal E}}{4\pi^2}\Bigl [\log
\Bigl (\frac {e{\cal E}}{\pi m^2}\Bigr )+1\Bigr ]-\frac {m^4}{16\pi}+m^4O
\Bigl (\frac {\pi m^2}{e {\cal E}}\Bigr )
\label{rate:approx}
\end {equation}
By comparison of eqs.(\ref {rate:exact}) and (\ref{rate:approx}), we get the 
estimate
\begin {equation} 
\sum_{n=1}^\infty \frac {1}{n^2}\e^{-nx}=\frac {\pi^2}{6}-x+x \log x-
\frac {x^2}{4}+O(x^3)
\label {est:series}
\end{equation}
that can be verified numerically, showing that eq.(\ref{rate:approx}) gives 
indeed the correct strong-field limit of the exact expression 
(\ref{rate:exact}).\par
Eq.(\ref{est:series}) is not a new result, as it merely follows from the 
correspondences used to get the minkowskian effective lagrangean. Now, with the 
aid of eq.(\ref{zpowm}), we will obtain the corrections to the rate $w$ due to 
the presence of a uniform magnetic field parallel to the electric one and much 
weaker than the latter. We find that the first correction (order of 
$({\cal B} /{\cal E})^2$) to the ${\cal B}=0$ effective lagrangean is
\begin {equation}
\delta^{(2)}{\cal L}_{\mathrm {Eff}}^{\mathrm M}({\cal E},{\cal B})
=\frac {e^2{\cal B}^2}{24\pi^2}
\biggl [
\psi \Bigl ({\mathrm i}\frac {m^2}{2e{\cal E}}\Bigr )+\log\Bigl (-{\mathrm i}
\frac{2e{\cal E}}{\mu^2}\Bigr )-{\mathrm i}\frac {e{\cal E}}{m^2}\biggr ]
\label {leff:b2}
\end {equation}
where $\psi(x)\equiv \d\log \Gamma (x)/\d x$.\par
An explicit form for the imaginary part of eq.(\ref {leff:b2}) can be achieved
by means of the inversion formula \cite {bateman}
for the $\psi$-function
\begin{equation}
\psi(z)-\psi(-z)= - \pi\, \cot \bigl (\pi z\bigr )-\frac {1}{z}
\label {psi}
\end {equation}
so we get 
\begin {equation}
{\mathrm {Im}}\ \delta^{(2)}{\cal L}_{\mathrm {Eff}}=\frac {e^2{\cal B}^2}{24\pi}
\,\frac {1}{\e^{(\pi m^2)/(e{\cal E})}-1}
\label {rate2}
\end {equation}
\par
From eqs.(\ref {rate:exact}) and (\ref {rate2}) we can compute exactly the 
total rate of production of fermion-antifermion pairs in a uniform 
electromagnetic field: we obtain (at $O({\cal B}^2/{\cal E}^2)$)
\begin {equation}
w=\frac {e^2{\cal E}^2}{4\pi^3}\ \sum_{n=1}^\infty\biggl [\,\frac {1}{n^2}+
\frac {\pi^2}{3}\, \frac {{\cal B}^2}{{\cal E}^2}\,\biggr ]\exp
\biggl \{-n\,\frac {\pi m^2}{e{\cal E}}\biggr \}+O\biggl (\frac {{\cal B}^4}
{{\cal E}^4}\biggr )
\label {rate:total}
\end {equation}
\par
It is worth noticing that neither eq.(\ref {rate:exact}) nor eq.(\ref {rate2})
are analytic in ${\cal E}=0$, owing to the non-perturbative character of this 
phenomenon.\par
According to a {\it na\"\i ve} interpretation, the phenomenon of particle 
production in external electromagnetic fields is due to the fact that
virtual pairs created (and, in the absence of the external field, 
annihilated) in the vacuum, are accelerated by the electric field and may gain 
energy enough to reach the threshold (i.e. the fermion mass) 
and become physical particles. Obviously, the more the 
electric field is strong, the more pairs are generated. 
On the other hand, electrically charged particles do not acquire
energy from a magnetic field, so that, in the light of the interpretation 
we have sketched above, the magnetic field should not give any contibution to 
the rate $w$. Thus, it is rather surprising that a (weak) magnetic field 
gives itself a contribution to the rate of pair creation, as we have shown in
eq. (25). Nevertheless, it is possible to check that the corrections we 
have obtained to the ${\cal B}=0$ case are the correct ones: considering also
the second term (of order $({\cal B}/{\cal E})^4$) to the 
"unperturbed" effective lagrangean and retaining only the first terms in the 
expansion for weak fields, we get the effective lagrangean in the limit
$|e{\cal B} |\ll|e{\cal E} |\ll m^2/2$:
\begin{equation}
\delta^{(4)}{\cal L}_{\mathrm {Eff}}^{\mathrm M}({\cal E},{\cal B})\simeq 
\frac {e^4}{16\pi^2}\
\frac {2}{45 m^4}\bigl [{\cal E}^4+{\cal B}^4+5{\cal E}^2{\cal B}^2\bigr ]+
\frac {e^2}{24\pi^2}{\cal B}^2\log
\frac {m^2}{\mu^2}
\label {eulheis}
\end{equation}
which is in perfect agreement with the Euler-Heisenberg effective lagrangean
\cite{EH}, after setting the arbitrary scale $\mu$ at the main scale $m$ of the 
problem.\par
A paper is recently appeared \cite{HH}, in which the effective lagrangean
of massive QED is obtained within the same approximations of this 
letter: the corrections to the $\pol{\cal E}\cdot\pol{\cal B}$ case shown in 
\cite{HH}
are (apart from some monomial terms coming from a different normalization) the 
same that can be derived from eq.(\ref {zintm:com}). The agreement with this
result confirms that the effective lagrangean consistently leads to the picture 
that the rate of pair production does really take a contribution from the 
magnetic component.\par
In conclusion, we notice that the effective lagrangean, when expressed as a 
power series of ${\cal E}/{\cal B}$, is real (in fact, 
only even 
powers of $y$ appear in eq.(\ref {zpowm})). 
As usual, anyway, we cannot exclude that an imaginary 
contribution, exponentially small when ${\cal E}\rightarrow 0$, indeed exists: 
according to the {\it na\"\i ve} interpretation we gave, such a contibution 
should really exist, even if we have not still been able to achieve a proof of 
its presence.

After having examinated the massive theory, taking special care to the 
rate of particle production, and bearing in mind that we could extract from 
eq.(\ref{zpowm}) the effective lagrangean for massive QED to any order in 
${\cal E}/{\cal B}$ or ${\cal B}/{\cal E}$ we turn our attention to the 
massless theory.

First, we would like to notice that the most characteristic feature of the 
massless case is encoded in the first term of the rhs of eq. (18), which 
appears to be an effect of infrared regularization from the $\zeta$-function 
technique.
It is easy to obtain, from eqs. (14) and (18), the effective 
euclidean lagrangean in the form
\begin {eqnarray}
{\cal L}_{\mathrm {Eff}}^{\mathrm E}=-\frac {e^2E^2}{8\pi^2}&\Bigl \{&
\frac {1}{3}\Bigl (1+\frac {B^2}{E^2}\Bigr )\log \frac {2|eE|}{\mu^2}\nonumber
\\
&+&4\zeta'_R(-1)-\frac {1}{3} -\frac {\gamma_E B^2}{3 E^2}+\nonumber\\
&-&\Bigl |\frac {B}{E}\Bigr |\log \frac {|eB|}{\pi\mu^2}\Bigr \}+
O\biggl (\Bigl (\frac {B}{E}\Bigr )^4\biggr )
\label {ventotto}
\end {eqnarray}
where the first and the third terms are the relevant ones, whereas the second 
one may always be subtracted after redefinition of the effective action 
up to polynomials in the electric and magnetic fields. The above expression
turns out to be real, which in fact suggests that the corresponding 
minkowskian quantity has to exhibit parity invariance. However, owing to 
our specific (and frame-dependent) choice of the field variables, an 
ambiguity concerning the Wick rotation seems to arise, due to the presence 
of the 
absolute value of $E$. This, however, is purely artificial because it is
always possible to choose $E$ positive, after suitable frame rotation -
we recall that our field configuration is that of parallel electric and
magnetic fields of different strengths.
Moreover, the very same Wick rotation actually
generates a nontrivial imaginary part in the Minkowskian effective action
(just like in the massive case - see eqs. (20) and (26)). The outcome 
that the corresponding pair creation rate is positive - i.e. the 
probability density to be within zero and one - indeed exhibits the
absence of any ambiguity. 

In particular, we find (up to polynomials in ${\cal E}$ and
${\cal B}$) 
\begin {eqnarray}
{\cal L}_{\mathrm Eff}^{\mathrm M}=\frac {\cal E^2}{2}-\frac {\cal B^2}{2}
&+& e^2\frac {{\cal E}^2-{\cal B}^2}{24 \pi^2}\Bigl \{\Bigl [1-\log
\frac {|2e{\cal E}|}{\mu^2}+i\frac {\pi}{2}\Bigr ]\nonumber \\ &+&i\frac
{e^2}{3}
|{\cal E}{\cal B}|\log \Bigl (\frac {|e{\cal B}|}{\pi\mu^2}\Bigr )\Bigr \}
+O\biggl (\Bigl (\frac {\cal B}{\cal E}\Bigr )^4\biggr )
\label {ventinove}
\end {eqnarray}
from which it follows, as expected, that the real part of the effective
lagrangean as well as the pair creation rate are both parity invariant.

To sum up, we have obtained some new expressions for the QED effective action
 in the 
presence of general uniform fields which extend the ones given in the 
literature, both 
in theories with massive and massless Fermi particles. Concerning the physical 
consequences, we focused our attention onto the rate of pairs production in a 
uniform electromagnetic field both in the massive and in the massless theory.
For these reasons we think
it is worth studying in a more complete way the effective theory of 
massless QED.

\end {document}